# Tremendous tunneling magnetoresistance effects based on van der Waals room-temperature ferromagnet $Fe_3GaTe_2$ with highly spin-polarized Fermi surfaces

Xinlu Li[1#], Meng Zhu[1#], Yaoyuan Wang[2], Fanxing Zheng[1], Jianting Dong[1], Ye Zhou[1], Long You[2], and Jia Zhang[1]*

[1]*School of Physics and Wuhan National High Magnetic Field Center,*

*Huazhong University of Science and Technology, 430074 Wuhan, China*`

[2]*School of Optical and Electronic Information,*

*Huazhong University of Science and Technology, 430074 Wuhan, China*

* jiazhang@hust.edu.cn;

[#]These authors contributed equally to this work.

**Abstract:**

Recently, van der Waals (vdW) magnetic heterostructures have received increasing research attention in spintronics. However, the lack of room-temperature magnetic order of vdW material has largely impedes its development in practical spintronics devices. Inspired by the recently discovered vdW ferromagnet $Fe_3GaTe_2$, which has been shown to have magnetic order above room temperature and sizable perpendicular magnetic anisotropy, we investigate the basic electronic structure and magnetic properties of $Fe_3GaTe_2$ as well as tunneling magnetoresistance effect in magnetic tunnel junctions (MTJs) with structure of $Fe_3GaTe_2$/Insulator/$Fe_3GaTe_2$ by using first-principles calculations. It is found that $Fe_3GaTe_2$ with highly spin-polarized Fermi surface ensures that such magnetic tunnel junctions may have prominent tunneling magnetoresistance effect at room temperature even comparable to existing conventional $AlO_x$ and MgO-based MTJs. Our results suggest that $Fe_3GaTe_2$-based MTJs may be the promising candidate for realizing long-waiting full magnetic vdW spintronic devices.

# I. Introduction

Since the discovery of long-range magnetic order down to the monolayer limit in $Cr_2Ge_2Te_6$[1] and $CrI_3$[2] materials in 2017, a wide variety of emerging two-dimensional (2D) magnetic materials have been discovered[3-6], which provide an effective materials platform for exploring novel 2D spintronics devices. Since then, great progress has been made both in theoretical and experimental studies on van der Waals (vdW) magnetic materials and related heterojunctions[7,8]. In particular, vdW heterojunctions with large magnetoresistance have shown potential applications in spintronics devices such as magnetic random-access memories and high-density hard disk drives.

For instance, $CrI_3$ has been used as a spin-filter barrier in graphite/$CrI_3$/graphite magnetic tunnel junctions (MTJs)[9,10], nevertheless, a large continuous magnetic field and extremely low temperature is required to maintain its magnetoresistance effect. Over the past few years, 2D magnetic heterojunctions using $Fe_3GeTe_2$ as ferromagnetic electrodes have been widely reported[11-17]. It has been demonstrated that tunneling magnetoresistance of 160% (at 4.2 K), 41% (at 10 K) and 300% (at 4.2 K) can be experimentally achieved in all vdW heterostructure $Fe_3GeTe_2$/$h$-BN/$Fe_3GeTe_2$[14], $Fe_3GeTe_2$/InSe/$Fe_3GeTe_2$[16] and $Fe_3GeTe_2$/h-BN/$Fe_3GeTe_2$[17], respectively. However, as the transition temperature of bulk $Fe_3GeTe_2$ (220 K)[4] are still below room temperature (300 K), a much lower operating temperature (< 10 K) is needed to achieve sizable TMR, which greatly hinders the practical applications of magnetic vdW heterostructures. As a result, it is crucial to search vdW heterostructures with both high transition temperatures and large TMR effect.

Recently, vdW ferromagnet $Fe_3GaTe_2$[18] with large perpendicular magnetocrystalline anisotropy (MCA) energy density (~4.79×10$^5$ J/m$^3$ at 300 K) and Curie temperature $T_C$ (350–380 K) well above room temperature has been discovered. Driven by this exciting finding, in this letter, we first examine the basic magnetic properties of $Fe_3GaTe_2$ and then investigate the spin-dependent electron transport in vdW MTJs based on $Fe_3GaTe_2$. In this case, $Fe_3GaTe_2$ is used as the ferromagnetic electrode of the MTJs, while for the spacer layer we have chosen insulator $h$-BN and semiconductor 2H-$WSe_2$. We demonstrate that the existence of a prevalent ultra-high tunneling magnetoresistance (TMR) effects, which mainly stem from the highly spin-polarized Fermi surfaces of $Fe_3GaTe_2$. Our findings may stimulate further experimental work on the realization of full vdW MTJs at room temperature.

## II. Calculation methods

By using Vienna ab initial Package (VASP)[19], we first examine either generalized gradient approximation (GGA-PBE)[20] or local density approximation (LDA)[21,22] exchange-correlation functions could better describe $Fe_3GaTe_2$ by comparing calculated magnetic properties to available experiments[18]. The calculated magnetic moment and magnetocrystalline anisotropy (MCA) energy are found to fit better with experimental values[18] by using LDA functionals than GGA as shown in Table 1. LDA will then be employed in the following calculation for $Fe_3GaTe_2$ in the rest of this work. The details of the calculations can be found in the Supplementary Information S1.

**Table 1**. The calculated average magnetic moment per Fe atom and MCA energy of bulk $Fe_3GaTe_2$ by using GGA and LDA functionals in comparison with experimental values. Please note that the quantities calculated by first-principles should be considered at 0 K.

| Magnetic properties of $Fe_3GaTe_2$ | GGA | LDA | Experimental values[18] |
|---|---|---|---|
| $M$ ($\mu_B$/Fe) | 1.93 | 1.76 | 1.68 (3 K) |
|  |  |  | 1.18 (300 K) |
| MCA energy density ($*10^5$ J/m$^3$) | 21.11 | 8.01 | 4.79 (300 K) |

The remaining first-principles calculations of this work are performed by Quantum Espresso simulation package[23] using the projector augmented wave (PAW) method with Garrity-Bennett-Rabe-Vanderbilt (GRBV) 1.5 ultrasoft pseudopotentials[24]. The LDA of Perdew-Zunger (PZ)[21] is used for the exchange-correlation functional potentials. We use the $15\times15\times3$ Monkhorst-Pack $k$-points mesh for the self-consistent calculation, while the energy cutoff for plane-wave and charge density are set at 40 and 320 Ry, respectively.

For the magnetic tunnel junction (MTJ) based on $Fe_3GaTe_2$, the self-consistently calculations of scattering region are performed by using a $k$-point mesh of 15×15×1. And then, the electron transmissions of the MTJs are obtained by using the wave-function scattering method[25], matching the wave function between the left and right $Fe_3GaTe_2$ electrodes and the scattering region of the MTJs. The $k_{//}=(k_x, k_y)$-resolved electron transmissions is then calculated with the 200×200 $k_{//}$ mesh in the 2D Brillouin zone (2DBZ). The spin-dependent ballistic conductance of the MTJs is calculated by summarizing transmission over entire 2DBZ based on the Landauer-Büttiker formula[26-28] as follow:

$$G_\sigma = \frac{e^2}{h}\sum_{k_{//}} T_\sigma(k_{//})$$

where $T_\sigma (k_{//})$ denotes the spin and $k_{//}$-resolved electron transmission, $e$ is the elementary charge, and $h$ is the Planck constant. The TMR ratio is defined as TMR = $(G_P - G_{AP})/G_{AP}$, where $G_P$ and $G_{AP}$ are the conductance for parallel (P) and antiparallel (AP) alignments of the magnetization orientations of two $Fe_3GaTe_2$ electrodes, respectively. The resistance-area product (*RA*) is another important physical quantity which can be evaluated from the conductance[13].

## III. Results and discussions

### Basic electronic structure and magnetic properties of $Fe_3GaTe_2$

The atomic structure of $Fe_3GaTe_2$ is the same as well studied $Fe_3GeTe_2$, which has a layered hexagonal crystal structure with a space group $P6_3/mmc$ (No. 194). The atomic structure of bulk $Fe_3GaTe_2$ is depicted in Fig. 1a, which illustrates the AB-stacking of $Fe_3GaTe_2$ with two-layer periodicity along the *c*-axis, separated by vdW gap. In each monolayer, the Fe atoms are located in two inequivalent Wyckoff sites $Fe^I$ and $Fe^{II}$ with different magnetic moments. The monolayer comprises five atomic layers with the central layer forming a $Fe^{II}$-Ga honeycomb lattice, then the next two layers form vertical $Fe^I$ dumbbells, while the top and bottom layers are occupied by Te sandwiching the $Fe_3Ga$ layers. The calculated spin-dependent band structure and density of states (DOS) of $Fe_3GaTe_2$ are shown in Fig. 1c and 1d. A large spin-polarization at the Fermi level can be observed from the band structure and DOS. From the LDA self-consistent calculations, the average magnetic moment of bulk $Fe_3GaTe_2$ is found to be 1.76 $\mu_B$/Fe, which agree well with the experimental value of 1.68 $\mu_B$/Fe[18] at 3 K as shown in Table 1.

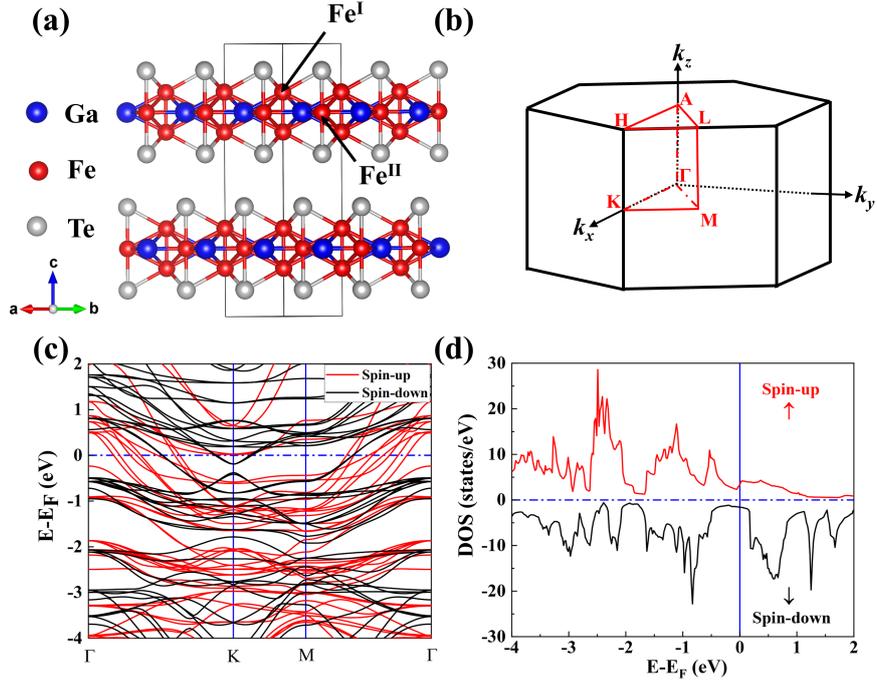

**Fig. 1.** (a) The crystal structure of bulk Fe$_3$GaTe$_2$, where the solid black box indicates the primitive cell of bulk Fe$_3$GaTe$_2$. (b) Brillouin zone and high-symmetry *k* path of a hexagonal lattice. (c) and (d) are the spin-polarized band structures and density of states for bulk Fe$_3$GaTe$_2$, respectively.

In the next, we first extract magnetic interaction parameters of Fe$_3$GaTe$_2$ and then perform Monte Carlo simulation to evaluate its Curie temperature. The spin Hamiltonian of Fe$_3$GaTe$_2$ can be modeled as follows:

$$H = -\sum_{i \neq j} J_{ij} S_i \cdot S_j - K \sum_i S_{i,z}^2$$

Where $J_{ij}$ is the Heisenberg exchange interaction between spin moment at site i and j, $S_i$ is a unit vector pointing to the local spin moment, $K$ is the uniaxial magnetic anisotropy constant. The interatomic Heisenberg exchange interactions have been calculated by using SPR-KKR package based on KKR-Green's function method using Lichtenstein's formula[29]. The calculated exchange interaction energy J$_{ij}$ for the first six nearest neighbors are listed in Table 2. Due to the large ferromagnetic exchange interaction for the first two nearest neighbors (J$_1$=57.18 meV, J$_2$=17.03 meV), Fe$_3$GaTe$_2$ is supposed to be ferromagnetic with high Curie temperature. The magnetic easy-axis has been determined by comparing the energy of Fe$_3$GaTe$_2$ with spin axis pointing along various directions by considering spin-orbit coupling (SOC) in calculations. It turns out that Fe$_3$GaTe$_2$ shows magnetic easy axis along *c*-axis. The calculated MCA energy of bulk Fe$_3$GaTe$_2$ evaluated from the magnetic force theorem is found to be around 1.116 meV/u.c corresponding to 8.01×10$^5$ J/m$^3$, which agrees with experimental magnetic anisotropy energy ($4.79 \times 10^5$ J/m$^3$ at 300 K).

Once the exchange interaction and magnetic anisotropy energy are extracted, the Curie temperature of Fe$_3$GaTe$_2$ is then determined by Monte Carlo simulations using UppASD package[30]. The Curie temperature $T_C$ of bulk and monolayer Fe$_3$GaTe$_2$ can be obtained by fitting the normalized magnetization $M/M_S$ by the following equation[31]

$$m(T) = \left(1 - \frac{T}{T_C}\right)^\beta$$

Where $\beta$ is the critical magnetization exponent with a typical value of around 0.3. As shown in Fig. 1b, the Curie temperature of bulk Fe$_3$GaTe$_2$ is fitted to be around 406 K, which is close to the experimental $T_C$ of 380 K[18]. The simulated Curie temperature of monolayer Fe$_3$GaTe$_2$ is 320 K, indicating that Fe$_3$GaTe$_2$ will remain ferromagnetic above room temperature even down to monolayer limit.

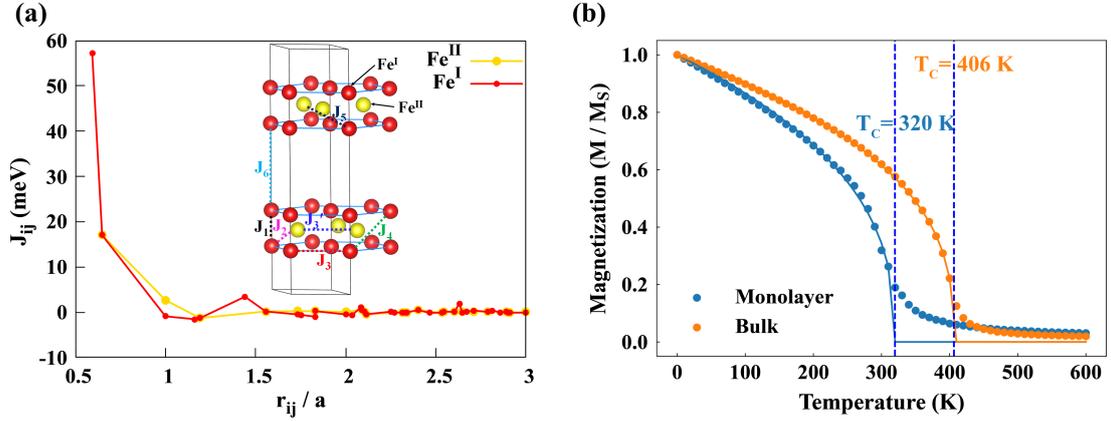

**Fig. 2.** (a)The calculated Heisenberg exchange interactions as a function of distance between pairs of Fe atoms by taking Fe$^I$ (red line) and Fe$^{II}$ (gold line) as center atoms in bulk Fe$_3$GaTe$_2$. The first six Fe-Fe exchange interaction path is indicated, where only the Fe atoms have been displayed in crystal structure. (b) The calculated normalized magnetization (depicted in solid circles) as a function of temperature for monolayer and bulk Fe$_3$GaTe$_2$ obtained from Monte Carlo simulation, where the solid lines are the fitted curve.

**Table 2.** The calculated Heisenberg exchange interaction $J_N$ of bulk Fe$_3$GaTe$_2$ for the first 6 nearest neighbors, where Ns is the coordination number.

| $J_N$ | Distance (Å) | Type | Ns | $J_{ij}$ (meV) |
|---|---|---|---|---|
| $J_1$ | 2.37 | Fe$^I$ - Fe$^I$ | 1 | 57.18 |
| $J_2$ | 2.59 | Fe$^I$ - Fe$^{II}$ | 6 | 17.02 |
| $J_3$ | 3.99 | Fe$^I$ - Fe$^I$ | 6 | -0.92 |
| $J_3'$ | 3.99 | Fe$^{II}$ - Fe$^{II}$ | 6 | 2.66 |
| $J_4$ | 4.64 | Fe$^I$ - Fe$^I$ | 6 | -1.55 |
| $J_5$ | 4.75 | Fe$^I$ - Fe$^{II}$ | 6 | -1.29 |
| $J_6$ | 5.74 | Fe$^I$ - Fe$^I$ (interlayer) | 1 | 3.43 |

## Tunneling magnetoresistance effect in the Fe$_3$GaTe$_2$-based vdW MTJs

In vdW magnetic tunnel junctions, spin-dependent transport and TMR at zero bias is mainly determined by the spin-polarization on Fermi surface of ferromagnetic electrode. Before investigating TMR effect in Fe$_3$GaTe$_2$-based vdW MTJs, we first calculate the Fermi surfaces of bulk Fe$_3$GaTe$_2$. Fig 3a and 3b show the spin-resolved three-dimensional (3D) Fermi surface of Fe$_3$GaTe$_2$ and Fig 3c and 3d illustrate its 2D Fermi surface projected onto 2DBZ. From the 2D Fermi surfaces, one can obtain the number of available Bloch states (conduction channels) of bulk Fe$_3$GaTe$_2$ at each $k_{//}=(k_x, k_y)$ points in the 2DBZ. It is clear that the 2D Fermi surface of spin down channel has few overlaps with spin up channel (Fig. 3e), which indicates highly spin-polarized Fermi surfaces, and should produce a large tunneling magnetoresistance effect in the vdW MTJs based on Fe$_3$GaTe$_2$.

For comparison, we also plot the 3D and 2D Fermi surfaces for other two well-studied vdW ferromagnets: Fe$_3$GeTe$_2$ (T$_C$~220 K)[4] and 1$T$-CrTe$_2$ (T$_C$~310 K)[32]. It can be observed that Fe$_3$GeTe$_2$ (Fig. 3f-g) have conduction channels around the Gamma point for both spin channels. As a result, the spin polarization on Fermi surface of Fe$_3$GeTe$_2$ is lower than Fe$_3$GaTe$_2$. On the other hand, although CrTe$_2$ have relatively high Curie temperature above 300 K, its two spin conduction channels have been overlapped in a large portion of BZ (Fig 3k-o), which indicates a low spin-polarization on Fermi surfaces and much lower TMR by using CrTe$_2$ as electrodes in vdW MTJs.

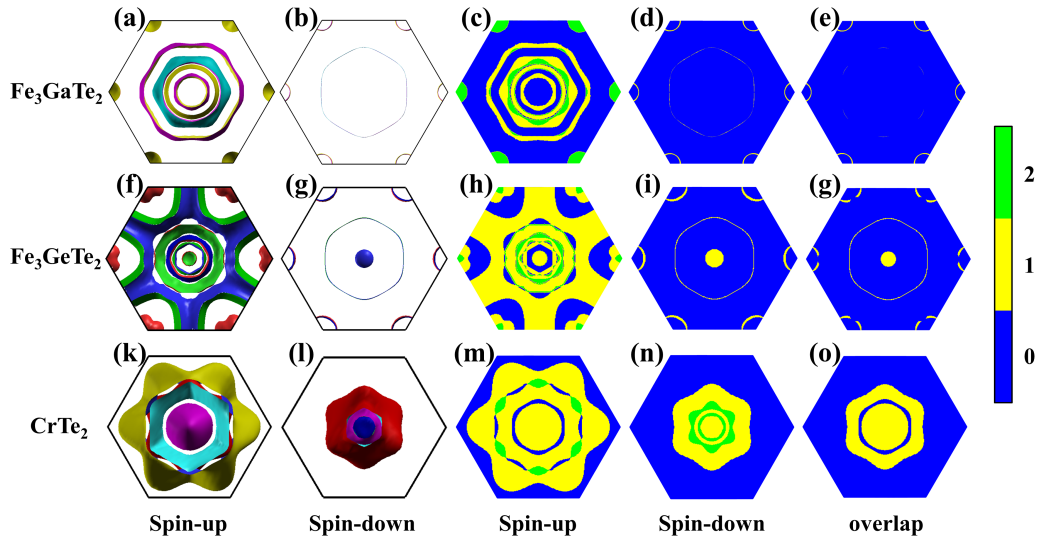

**Fig. 3.** (a) and (b) are the spin-up and spin-down 3D Fermi surfaces of bulk Fe$_3$GaTe$_2$, respectively, which are plotted using the xcrysden package[22]. (c) and (d) are the projected Fermi surfaces in the 2DBZ of bulk Fe$_3$GaTe$_2$. (e) shows the overlapped region for spin up and spin down channels. (f) - (g) and (k) - (o) show the corresponding 3D, 2D and overlapped Fermi surfaces for bulk Fe$_3$GeTe$_2$ and CrTe$_2$, respectively. The color bar at the right side of lower

panel indicates the number of available Bloch states at Fermi level.

Based on the above analysis, we build vdW MTJs comprised of $Fe_3GaTe_2$ electrodes and monolayer *h*-BN barrier, using the $\sqrt{3} \times \sqrt{3}$ in-plane unit cell of *h*-BN matched at the interface with the unit cell of $Fe_3GaTe_2$ electrode, as illustrated in Fig. 4a. The optimized vertical interlayer space between $Fe_3GaTe_2$ electrodes and *h*-BN barrier is 3.50 Å, which is a typical vdW interaction distance. We then calculate the tunneling conductance of the MTJs for parallel (Fig. 4b and 4c) and antiparallel (Fig. 4d) magnetization alignments of the left and right $Fe_3GaTe_2$ electrodes. The spin-up (Fig. 4b) and spin-down (Fig. 4c) electron transmission for parallel configuration are in accordance with the conduction channels of bulk $Fe_3GaTe_2$ shown in Fig. 3c and 3d. For antiparallel configuration, the electron transmission shown in Fig. 4d is consistent with the overlapped Fermi surface shown in Fig.3e and have been greatly suppressed owing to the mismatch of Fermi surfaces for two spin channels. In consequence, a tremendous TMR over 30000% has been achieved in $Fe_3GaTe_2$/*h*-BN/$Fe_3GaTe_2$ MTJ.

Without loss of generality, we also build another $Fe_3GaTe_2$-based MTJs by considering semiconductor 2H-$WSe_2$ and vacuum as tunnel barriers (Supporting Information Note 2). As listed in Table 3, $Fe_3GaTe_2$-based MTJs with both tunnel barriers all exhibit similarly large TMR (>10000%). It is worthy pointing out that such large TMR of $Fe_3GaTe_2$-based MTJs is around one order of magnitude larger than previously studied $Fe_3GeTe_2$-based MTJs[11]. Furthermore, we also compare the transport properties of $Fe_3GaTe_2$-based MTJs to another room temperature ferromagnet $CrTe_2$-based MTJs. By using insulator $\beta$-$MgCl_2$[33] (band-gap 5.42 eV) as tunnel barrier, the $CrTe_2$/$MgCl_2$/$CrTe_2$ MTJs are illustrated in Fig. 4b. As it can be seen from the spin-dependent transmission of $CrTe_2$-based MTJs shown in Fig. 4f-4h, the small mismatch between spin up and spin down conduction channels results in large transmission for antiparallel alignment in 2DBZ, which turns out to produce much smaller TMR of only around 25% (Table 3). These calculation results clearly indicate that $Fe_3GaTe_2$-based MTJs show great superiority over the existing vdW MTJs.

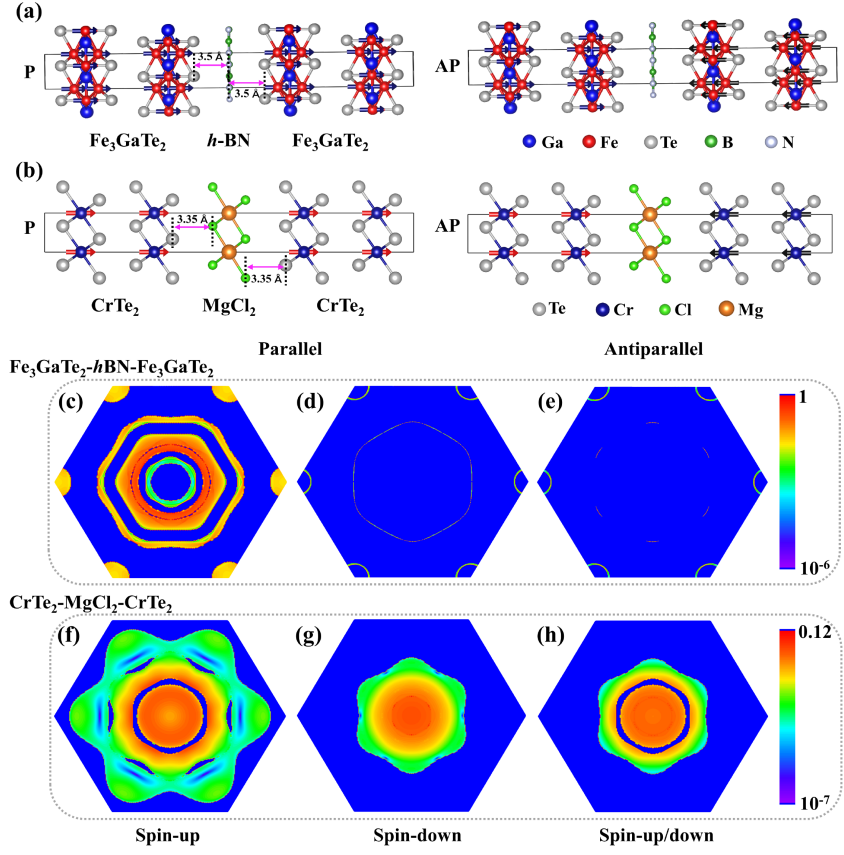

**Fig. 4.** Atomic structures of Fe$_3$GaTe$_2$/h-BN/Fe$_3$GaTe$_2$ (a) and CrTe$_2$/MgCl$_2$/CrTe$_2$ (b) MTJs with parallel (P) and antiparallel (AP) magnetization alignments of the two electrodes. Arrows on Fe and Cr atoms denote the direction of magnetic moments. (c) and (d) are the spin-up and spin-down $k_{//}$-resolved transmission distributions in 2DBZ of Fe$_3$GaTe$_2$/h-BN/Fe$_3$GaTe$_2$ in parallel configuration, respectively. (e) is the either-spin transmission for antiparallel state. (f)-(h) show the corresponding transmission of CrTe$_2$/MgCl$_2$/CrTe$_2$. Color scale alongside indicates the available transmission on a logarithmic scale.

**Table 3.** The spin-dependent transport properties of several different vdW MTJs when the magnetizations of two ferromagnetic electrodes are in P and AP alignments.

| MTJ structures | Transmission of P state | | Transmission of AP state | | TMR (%) |
|---|---|---|---|---|---|
| | $T$ | $RA$ ($\Omega \times \mu m^2$) | $T$ | $RA$ ($\Omega \times \mu m^2$) | |
| Fe$_3$GaTe$_2$/h-BN/Fe$_3$GaTe$_2$ | $1.77 \times 10^{-2}$ | **0.20** | $5.49 \times 10^{-5}$ | **64.66** | $3.21 \times 10^4$ |
| Fe$_3$GaTe$_2$/2H-WSe$_2$/Fe$_3$GaTe$_2$ | $9.79 \times 10^{-2}$ | **0.11** | $8.78 \times 10^{-4}$ | **12.13** | $1.11 \times 10^4$ |
| Fe$_3$GaTe$_2$/vacuum/Fe$_3$GaTe$_2$ | $5.89 \times 10^{-5}$ | **60.27** | $2.81 \times 10^{-8}$ | **1.26×10$^5$** | $2.10 \times 10^5$ |
| Fe$_3$GeTe$_2$/h-BN/Fe$_3$GeTe$_2$[11] | $2.15 \times 10^{-2}$ | **0.17** | $3.38 \times 10^{-4}$ | **10.53** | $6.26 \times 10^3$ |
| Fe$_3$GeTe$_2$/vacuum/Fe$_3$GeTe$_2$[11] | $6.18 \times 10^{-3}$ | **0.58** | $4.44 \times 10^{-5}$ | **80.16** | $1.38 \times 10^4$ |
| CrTe$_2$/MgCl$_2$/CrTe$_2$ | $3.81 \times 10^{-3}$ | **0.84** | $3.04 \times 10^{-3}$ | **1.06** | 25.12 |

The aforementioned transport investigations under zero bias of $Fe_3GaTe_2$-based MTJs are all conducted at Fermi energy ($E_F$). In reality, the Fermi level may shift away from $E_F$ caused by doping, lattice strain, structural defects, *etc*. To further clarify the robust of large TMR in $Fe_3GaTe_2$-based MTJs, we investigate the energy dependence of Fermi surfaces for bulk $Fe_3GaTe_2$ within the energy range between $E_F$-0.15 eV and $E_F$+0.15 eV. As depicted in Fig. 5, in the studied energy range, the highly spin-polarized Fermi surfaces of $Fe_3GaTe_2$ have been preserved. Noteworthily, at $E_F$+0.1 eV and $E_F$+0.15 eV, the conduction channels between spin-up and spin-down are completely mismatched, corresponding to fully spin-polarized Fermi surfaces. By taking $Fe_3GaTe_2$/*h*-BN/$Fe_3GaTe_2$ MTJ as an example, as illustrated in Fig. 6, we explicitly calculate transmission and TMR as a function of energy. It turns out that for energy below $E_F$, the TMR ratios tend to first increase and then decrease as the energy decreases. Whereas for energy above $E_F$, the TMR ratios increase as the energies increases and reach infinity when energy is larger than 0.1 eV above $E_F$, due to the 100% spin-polarization of Fermi surfaces in 2DBZ. This result indicates that the $Fe_3GaTe_2$-based MTJ should have ultra-high TMR ratios in relatively wide energy window around Fermi energy.

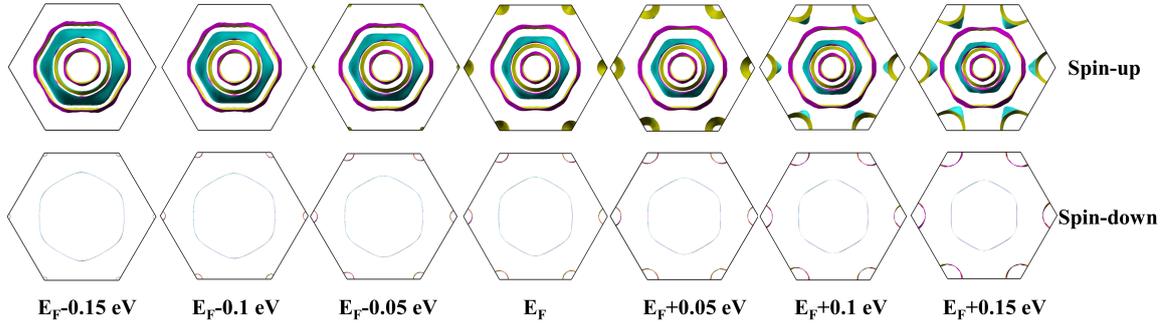

**Fig. 5.** The spin-up (upper panel) and spin-down (lower panel) Fermi surfaces of bulk $Fe_3GaTe_2$ around Fermi energy ranging from $E_F$-0.15 to $E_F$+0.15 eV.

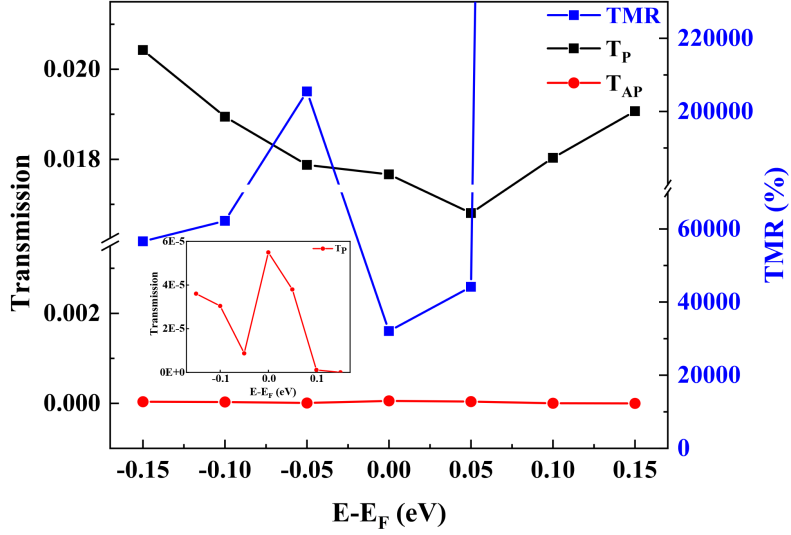

**Fig. 6.** The electron transmission for the parallel ($T_P$, in black) and antiparallel ($T_{AP}$, in red) states and the TMR (in blue, refer to right axis) as a function of energy in the Fe$_3$GaTe$_2$/$h$-BN/Fe$_3$GaTe$_2$ MTJ. The electron transmission for the parallel ($T_P$, in black) and antiparallel ($T_{AP}$, in red) states and the TMR (in blue, refer to right axis) as a function of energy in the Fe$_3$GaTe$_2$/$h$-BN/Fe$_3$GaTe$_2$ MTJ. The inset shows an enlarged view of transmission for antiparallel state as a function of energy.

## IV. Summary and Conclusion

In summary, the spin-dependent transport properties in vdW MTJs based on ferromagnet Fe$_3$GaTe$_2$ have been investigated by first-principles calculations. Monte Carlo simulations have shown that the bulk Fe$_3$GaTe$_2$ have high Curie temperature above room temperature. The Fe$_3$GaTe$_2$-based vdW MTJs have an ultra-high TMR effect over 10000% originating from the highly spin-polarized Fermi surfaces. Moreover, the huge TMR effect is robust against the variation of Fermi energy. Our results indicate that Fe$_3$GaTe$_2$-based vdW MTJs are very promising for realizing remarkable room temperature TMR effect, which may be comparable or even larger than traditional CoFe/AlO$_x$/CoFe and CoFeB/MgO/CoFeB MTJs. Our calculations provide a theoretical insight that may stimulate further experimental investigations on Fe$_3$GaTe$_2$-based magnetic vdW junctions towards high-performance room temperature spintronics devices.

## ACKNOWLEDGMENTS

J. Z. is supported by the National Natural Science Foundation of China with grant No. 12174129.

*jiazhang@hust.edu.cn;